\documentclass[12pt]{article}
\usepackage{amssymb}
\usepackage{amsmath, epsfig, rotating}
\usepackage{comment, color}

\usepackage{mathtools}
    \mathtoolsset{centercolon}

\usepackage{bm}

    \vfuzz \hfuzz
\renewcommand{\vec}[1]{\boldsymbol{#1}}

\newcommand{\tr}{\ensuremath{\operatorname{tr}}}

  \providecommand*{\pder}[2]{\ensuremath{\frac{\partial #1}{\partial #2}}}

\begin{document}

\title{Proper formulation of viscous dissipation \\ for nonlinear waves in solids}



\author{Michel Destrade$^a$, Giuseppe Saccomandi$^b$, Maurizio Vianello$^c$\\[6pt]
$^a$School of Mathematics, Statistics and Applied Mathematics, \\
National University of Ireland Galway, Ireland.\\[6pt]
$^b$Dipartimento di Ingegneria Industriale, \\
Universit\`{a} degli Studi di Perugia, 06125 Perugia, Italy. \\[6pt]
$^c$Dipartimento di Matematica, \\
Politecnico di Milano, \\
Piazza Leonardo da Vinci 32, I-20133 Milano, Italy.}

\date{}
\maketitle

\begin{abstract}

In order to model nonlinear viscous dissipative motions in solids, acoustical physicists usually add terms linear in $\vec{\dot{E}}$, the material time derivative of the Lagrangian strain tensor $\vec E$, to the elastic stress tensor  $\vec\sigma$ derived from the expansion to the third- (sometimes fourth-) order of the strain energy density $\mathcal E=\mathcal E(\tr\vec E, \tr\vec E^2, \tr\vec E^3)$. Here, it is shown  that this practice, which has been widely used in the past three decades or so, is physically wrong for at least two reasons, and that it should be corrected. One reason is that the elastic stress tensor $\vec \sigma$ is not symmetric while $\vec{\dot{E}}$ is symmetric, so that motions for which $\vec \sigma + \vec \sigma^T \ne \vec 0$ will give rise to elastic stresses which have no viscous pendant. Another reason is that $\vec{\dot{E}}$ is frame-invariant, while $\vec \sigma$ is not, so that an observer transformation would alter the elastic part of the total stress differently than it would alter the dissipative part, thereby violating the fundamental principle of material frame-indifference.  These problems can have serious consequences for nonlinear shear wave propagation in soft solids, as seen here with an example \color{black} of a kink in almost incompressible \color{black}  soft solids.

\end{abstract}

\newpage


\section{Introduction and main statement}


Nonlinear elastic wave propagation is a subject of considerable interest for many scientific and industrial applications such as geophysical exploration, soft tissue acoustics, and the dynamics of rubbers, silicones, and gels. From a theoretical point of view the mathematics and mechanics of nonlinear wave phenomena is a classical yet still active subject of research where many outstanding problems are awaiting a definitive systematic treatment.

In Physical Acoustics, the expansion of the strain energy density $\mathcal E$ to include nonlinear corrections is often attributed to Landau. Hence at the ``third-order" in the Lagrangian strain tensor $\vec E$, we write
\begin{equation}
    \mathcal E = \mu I_2 + \tfrac{1}{2}\lambda I_1^2 +
    \tfrac{1}{3}A I_3 + B I_1 I_2 + \tfrac{1}{3}C I_1^3,
\end{equation}
where $I_k=\tr\vec E^k$ ($k=1,2,3$) are the strain invariants, $\lambda$ and $\mu$ are the (second-order) Lam\'e coefficients, and $A$, $B$, $C$ are the (third-order) Landau coefficients. From an historical point of view, it is interesting to note that this notation by Landau seems to have appeared first in the \emph{Theory of Elasticity} (1986), co-written with Lifshitz, as an unnumbered equation in an exercise\cite{Landau1986}. In 1937, Landau\cite{Landau37} had already provided the third-order expansion of $\mathcal E$, that time denoting the third-order elastic constants by the letters  $P'$, $Q'$, $R'$. That same year, Murnaghan\cite{Murnaghan37} also proposed a third-order expansion using different strain invariants, which is still in use today (mostly by geophysicists). However the paternity of the third-order expansion can be traced further back in time, at least to a 1925 paper by Brillouin\cite{Brillouin25}, who in fact seems to have also been the first to use the letters $A$, $B$, $C$ for third-order elastic constants.

In any event, Landau and Lifshitz denote by $x_k$ the Lagrangian coordinate components, by $u_k$ the mechanical displacement components, and they derive the equations of motion in Cartesian coordinates as follows,
\begin{equation}\label{motion}
    \dfrac{\partial \sigma_{ik}}{\partial x_k} = \rho_0 \dfrac{\partial^2
    u_i}{\partial t^2},
\end{equation}
where $\rho_0$ is the mass density in the undeformed configuration, and the stress tensor is related to the strain through
\begin{equation}\label{stress}
    \sigma_{ik} = \dfrac{\partial \mathcal E}{\partial\left(\partial u_i/\partial
    x_k\right)}.
\end{equation}
(It is easy to recognize that this is the ``first Piola-Kirchhoff stress tensor'' of Classical Continuum Mechanics.) For the reader's convenience, we recall that the full Lagrangian strain tensor $\vec E$ has Cartesian components
\begin{equation}\label{lagrange}
    E_{ik} =
    \frac{1}{2}
    \left(
    \dfrac{\partial u_i}{\partial x_k} + \dfrac{\partial u_k}{\partial x_i}
    +\dfrac{\partial u_j}{\partial x_i} \dfrac{\partial u_j}{\partial x_k}
    \right),
\end{equation}
and can be rewritten as
\begin{equation}\label{lagrangebis}
    E_{ik} = \tfrac{1}{2}\left(F_{ji}F_{jk}-\delta_{ik}\right).
\end{equation}
Here, $F_{ik}$ are the Cartesian components of $\vec{F}$, the deformation gradient, defined as
\begin{equation}\label{definition.F}
    F_{ik} = \pder{y_i}{x_k}=\delta_{ik} + \dfrac{\partial u_i}{\partial x_k} \ ,
\end{equation}
where $y_i(x_k)$ are the space coordinates of the current position of material point $x_k$, for the deformation given by $y_i(x_k)=x_i+u_i(x_k)$.

In another part of their book (more precisely: at the end of \S 34 in Chapter 5) Landau and Lifchitz model \emph{viscous dissipation} by adding to the elastic stress tensor $\vec\sigma$ a ``viscosity stress tensor'' $\vec{\sigma'}$, with Cartesian components
\begin{equation}\label{viscous}
\begin{aligned}
    \sigma'_{ik} &=2\eta(\dot{E}_{ik}-\tfrac{1}{3}\delta_{ik}\dot{E}_{ll})
    + \zeta\dot{E}_{ll}\delta_{ik} \\
    &=  (\zeta - \tfrac{2}{3}\eta) \, \dot E_{ll} \delta_{ik} + 2\eta \,
    \dot E_{ik}\,,
\end{aligned}
\end{equation}
where $\eta>0$ and $\zeta>0$ are the shear and bulk viscosity coefficients, respectively, and the superposed dot denotes the time derivative. Note that it is completely unambiguous from the context of this part of their book that Landau and Lifshitz are speaking here of the \emph{infinitesimal} theory of visco-elasticity, and that $\vec{\dot{E}}$ in Eq.~\eqref{viscous} is the time derivative of the infinitesimal strain tensor (i.e. the linear part of  Eq.~\eqref{lagrange}), see their unnumbered equation between (34.2) and (34.3). It might seem at first glance that adding $\vec{\sigma'}$ to $\vec \sigma$, and taking the strain components to be those of the finite Lagrangian strain \eqref{lagrange} instead of the infinitesimal strain, would be a first, logical step towards the inclusion of nonlinear dissipative effects. However there are at least two problems with this seemingly anodyne approach.

One problem is that $\vec{\sigma'}$ is symmetric while $\vec{\sigma}$ is not. Then the resultant total stress tensor $\vec \sigma + \vec{\sigma'}$ has a visco-elastic symmetric part: $(\vec \sigma + \vec \sigma^T)/2 + \vec{\sigma'}$, but a purely elastic anti-symmetric part: $(\vec \sigma - \vec \sigma^T)/2$, where the superscript $T$ denotes the transpose. It is thus impossible to model antisymmetric viscous stress effects in motions and boundary conditions with this formulation.

The other problem is that $\vec{\sigma}$ and $\vec{\sigma'}$ are made \emph{objective}, or \emph{frame-indifferent}, in two different and irreconcilable ways. Indeed, \color{black}it is well known \color{black} that there are transformation rules to follow in order to ensure that the directions associated with a tensor are unaltered by an observer transformation. To summarize, two observers, one associated with a frame with position $\vec x$ and time $t$ and the other associated with a frame with position $\vec x^*$ and time $t^*$ are said to be equivalent when they are connected by (see e.g. Chadwick\cite{Chad99}),
\begin{equation}
    \vec x^* = \vec c(t) + \vec Q(t)\vec x, \qquad t^* = t-a,
\end{equation}
where $\vec Q$ is a proper orthogonal tensor, $\vec c$ is a vector, and $a$ is a constant scalar. The transformation rule for the elastic part of the stress tensor \eqref{stress} is that $\vec \sigma$ transforms into $\vec{\overline{\sigma}}$ given by
\begin{equation}
    \vec{\overline{\sigma}} = \vec{Q \sigma},
\end{equation}
in order to ensure frame-indifference, whilst for the viscous part \eqref{viscous}, it is
\begin{equation}
    \vec{\overline{\sigma'}} = \vec{\sigma'},
\end{equation}
because $\vec E$ -- and hence $\vec{\dot{E}}$ -- is observer-invariant (this is not so obvious from Eq.~\eqref{lagrange}, but becomes more so in view of Eq.~\eqref{lagrangebis}, as shown in Ref.~\cite{Chad99}). Clearly, it is not possible for the composite stress tensor $\vec \sigma + \vec{\sigma'}$ to comply with both requirements simultaneously and appear the same to two equivalent observers. This state of affair violates the fundamental principle of objectivity, a cornerstone of rational mechanics, e.g. see Gurtin\cite{Gur81}.

Now, there are two ways to reconcile $\vec \sigma$ and $\vec{\sigma'}$ in order to construct a stress tensor coherent with respect to symmetry and objectivity. One way is to make $\vec \sigma$ behave like $\vec{\sigma'}$, by replacing it with $ \vec F^{-1} \vec \sigma$, the (symmetric) second Piola-Kirchhoff stress tensor; the other way is to make $\vec{\sigma'}$ behave like $\vec \sigma$, by replacing it with $\vec{{\bar\sigma'}} = \vec{F\sigma'}$.

Evidently, both courses of action are eventually equivalent. In weakly nonlinear elasticity theory, governing equations have been around for $\vec \sigma$ longer than for $\vec{\sigma'}$ and we henceforth take advantage of these and concentrate on the consequences of modifying $\vec{\sigma'}$ rather than $\vec \sigma$. We thus propose the following form for the viscous stress, $\vec{{\bar{\sigma}'}} = \vec{F\sigma'}$, with components
\begin{equation}\label{viscous2}
\begin{aligned}
    {\bar{\sigma}'_{ik}} &= F_{ij}\sigma'_{jk} \\
    &= \left(\zeta - \tfrac{2}{3} \eta \right)F_{ik}\, \dot E_{ll} +
    2\eta F_{ij}\,\dot E_{jk} \\
   &= \left(\zeta - \tfrac{2}{3} \eta \right)\left(\delta_{ik} +
    \dfrac{\partial u_i}{\partial x_k} \right)\, \dot E_{ll} +
    2\eta\left(\delta_{ij}
    + \dfrac{\partial u_i}{\partial x_j}\right) \, \dot E_{jk},
\end{aligned}
\end{equation}
and investigate the consequences of using $\vec{\bar{\sigma}'}$ instead of $\vec{\sigma'}$ in the equations of motion.

In passing, it is interesting to show the form taken by the constitutive equation for the \emph{Cauchy stress tensor} $\bar{\vec\tau}'$, related to $\bar{\vec\sigma}'$ by
\begin{equation}\label{relation.Piola.Cauchy}
  j \bar{\tau}'_{ik}=\bar{\sigma}'_{is}F_{ks},
\end{equation}
\color{black}
where $j=\det\vec F$.
For this we recall that\cite{Chad99}
\color{black}
\begin{equation}\label{dot.u.ik}
 \dot{E}_{ik} =  F_{li} d_{lj}F_{jk},
\end{equation}
where $d_{lj}$ is the Cartesian component of $\vec d$, the Eulerian stretching tensor,
\begin{equation}\label{def.d.lj}
  d_{lj}=\frac12\left(\pder{v_l}{y_j}+\pder{v_j}{y_l}\right).
\end{equation}
and that the (symmetric) left Cauchy-Green strain tensor $\vec b$ is defined as
\begin{equation}
  b_{jl}=F_{ji}F_{li},
\end{equation}
so that, in view of Eq.~\eqref{dot.u.ik}, the expression for $\dot{E}_{ii}$ can be written as
\begin{equation}\label{dot.u.ii}
    \dot{E}_{ii}= F_{li} d_{lj}F_{ji}=b_{jl}d_{lj}=\tr(\vec b\vec d).
\end{equation}
Thus, Eq.~\eqref{viscous2} yields
\begin{equation}
\begin{aligned}
  \bar{\sigma}'_{ik} &=  \left(\zeta - \tfrac{2}{3} \eta \right)F_{ik}\,\dot E_{ll}
  + 2\eta F_{ij}\,\dot E_{jk} \\
    &= \left(\zeta - \tfrac{2}{3} \eta \right)F_{ik}\,b_{jl}d_{lj} +
    2\eta F_{ij}\,F_{lj} d_{ls}F_{sk} \\
    &= \left(\zeta - \tfrac{2}{3} \eta \right)F_{ik}\,b_{jl}d_{lj} +
    2\eta b_{il} d_{ls}F_{sk}.
\end{aligned}
\end{equation}
From Eqs.~\eqref{relation.Piola.Cauchy}, \eqref{dot.u.ik} and \eqref{dot.u.ii} we compute the viscous part $\bar{\tau}'_{ik}$ of the Cauchy stress as
\begin{equation}
\begin{aligned}
  j\bar{\tau}'_{im} &= \bar{\sigma}'_{ik}F_{mk} \\
   &= \left(\zeta - \tfrac{2}{3} \eta \right)F_{ik}F_{mk} b_{lj} d_{jl} +
   2\eta b_{il} d_{ls} F_{sk} F_{mk} \\
   &= \left(\zeta - \tfrac{2}{3} \eta \right) b_{im} b_{il} d_{ls}
   + 2\eta b_{il}d_{ls}b_{sm},
\end{aligned}
\end{equation}
or, in absolute notation,
\begin{equation}\label{expression.for.bar.tau.prime}
  j\bar{\vec{\tau}}'= \left(\zeta - \tfrac{2}{3} \eta \right) \vec b \tr(\vec b\vec d)
   + 2\eta \vec b\vec d\vec b.
\end{equation}
It is quite clear that this is a polynomial isotropic function of the two tensor variables $\vec b$ and $\vec d$, which can be shown to be a special case of the universal representation for such functions (see, \textit{e.g.}, the classical treatise by Truesdell and Noll \cite[Sect.~13]{TN65} and the paper by Rivlin\cite{Rivlin1955}).
It is reassuring to notice that expression \eqref{expression.for.bar.tau.prime} for the viscous part of Cauchy stress $\bar{\vec{\tau}}'$ falls perfectly within the class of constitutive equations for finite viscoelasticity. This makes much more plausible our proposal \eqref{viscous2} for a modified version of  Eq.~\eqref{viscous}.


\section{Wave motion}
\label{section2}


When it comes to study wave propagation, attention is usually focused on special motions, for which the general equations of motion simplify greatly. One might then be led to believe that no relevant corrections should be made on the governing equations as a consequence of the substitution of Eq.~\eqref{viscous} with the more appropriate version Eq.~\eqref{viscous2}. In this section we show that this is \emph{not} the case, indeed, and that some relevant and quite complicated changes should be made to the propagation equations, even for simple (bulk) motions.

We discuss the propagation in a soft material of a transverse wave described by a displacement field $u_i(x_k,t)$ of the form
\begin{equation}\label{wave.motion}
    u_1=u(z,t), \qquad u_2=u_3=0,
\end{equation}
where $z$ is the third Cartesian coordinate. This is the choice made by Zabolotskaya et al.\cite{Zabolotskaya2004}, who added a viscous part in the form \eqref{viscous} to an elastic constitutive equation (see their Eq.~34).

Straightforward computations show that
\begin{equation} 
    \vec F=
    \begin{bmatrix}
      1 & 0 & u_z \\
      0 & 1 & 0 \\
      0 & 0 & 1 
    \end{bmatrix},
    \qquad
  \vec E= \dfrac{1}{2}
  \begin{bmatrix}
    0 & 0 &  u_z \\
    0 & 0 & 0 \\
    u_z & 0 & u_z^2 
  \end{bmatrix},
\end{equation}
(here, for compactness, we denote partial derivatives with subscripts, while in other formulas we use the more explicit notation, so that a comparison with some relevant references is made easier).

The time derivative of $\vec E$ and its trace are thus given by
\begin{equation}
  \dot{\vec E}=
  \dfrac{1}{2}\begin{bmatrix}
    0 & 0 &  u_{zt} \\
    0 & 0 & 0 \\
     u_{zt} & 0 & 2 u_z u_{zt} \\
  \end{bmatrix},
  \qquad
  \text{tr}(\dot{\vec{E}})= u_z u_{zt} ,
\end{equation}
and, moreover,
\begin{align}
  \vec F\dot{\vec E} =
  &  \dfrac{1}{2}
    \begin{bmatrix}
      1 & 0 & u_z  \\
      0 & 1 & 0 \\
      0 & 0 & 1 \\
    \end{bmatrix}
\begin{bmatrix}
    0 & 0 &  u_{zt} \\
    0 & 0 & 0 \\
    u_{zt} & 0 & 2 u_z u_{zt} \\
  \end{bmatrix}\notag \\[8pt]
   =& 
  \dfrac{1}{2}
  \begin{bmatrix}
  u_z u_{zt} & 0 &
  u_{zt} + 2 u_z^2 u_{zt} \\
    0 & 0 & 0 \\
   u_{zt}  & 0 & 2 u_z u_{zt} \\
  \end{bmatrix}
\end{align}

Now, since by Eq.~\eqref{viscous2}
\begin{equation}
    {\vec{\bar{\sigma}}}' = (\zeta-\tfrac{2}{3}\eta)\tr(\dot{\vec{E}})\,\vec F
    +2\eta\vec F\dot{\vec E},
\end{equation}
we find that the Cartesian components of the viscous part ${\bar{\vec\sigma}'}$ of the Piola-Kirchhoff stress tensor are given by
\begin{equation}
  \begin{bmatrix}
    (\zeta+\frac13\eta)u_z u_{zt} & 0 & \eta u_{zt}+(\zeta +\frac43 \eta) u_z^2 u_{zt} \\
    0 & (\zeta-\frac23 \eta) u_z u_{zt} & 0 \\
    \eta u_{zt} & 0 & (\zeta+\frac43 \eta) u_z u_{zt} 
  \end{bmatrix}.
\end{equation}

In their Section III,  Zabolotskaya et al.\cite{Zabolotskaya2004} investigate the propagation of waves described by Eq.~\eqref{wave.motion} in \color{black} an almost incompressible \color{black} soft material with  elastic strain energy
\begin{equation}\label{ZH.Energy}
    \mathcal E = \mu I_2 +\tfrac{1}{3}A I_3 + D I_2^2,
\end{equation}
where $\mu$, $A$, $D$ are second-, third-, and fourth-order elastic constants, respectively. Terms of order higher than the third in the strain ${\partial u}/{\partial z}$ are then neglected and the governing equations reduce to the single equation
\begin{equation}\label{equation.motion.ZH}
  \rho_0 \dfrac{\partial^2 u}{\partial t^2}=\mu \dfrac{\partial^2 u}{\partial z^2}+\gamma \dfrac{\partial}{\partial z}\left(\dfrac{\partial u}{\partial z}\right)^3,
\end{equation}
where $\rho_0$ is the constant density and $  \gamma=\mu+ \tfrac{1}{2}A +D$, see Ref.~\cite[Eq.~12]{Zabolotskaya2004}.

Next, in Section V of  Ref.\cite{Zabolotskaya2004}, a viscous stress is added, in the form of $\sigma'_{ik}$ as defined here by Eq.~\eqref{viscous}, and an appropriate modification for the governing equation \eqref{equation.motion.ZH} is deduced, which leads to the introduction of an additional viscous term:
\begin{equation}\label{viscous.ZH}
   \rho_0 \dfrac{\partial^2 u}{\partial t^2} =
   \underbrace{\mu  \dfrac{\partial^2 u}{\partial z^2}+\gamma \dfrac{\partial}{\partial z}\left(\dfrac{\partial u}{\partial z}\right)^3}_{\text{elastic}}
  + \underbrace{\eta  \dfrac{\partial^3 u}{\partial z^2 \partial t}}_{\text{viscous}}.
\end{equation}
This equation can also be found in the Refs.\cite{Woch08, Renier08, Jacob07}, for instance.
Our aim here is to verify which modifications would be required for the wave equation \eqref{equation.motion.ZH} if the added viscous Piola-Kirchhoff stress were defined by Eq.~\eqref{viscous2}, as we suggest, rather than by Eq.~\eqref{viscous}. In other words: what difference does it make to the wave propagation equations to consider such a stress tensor ${\bar{\sigma}}'_{ik}$, rather than ${\sigma}'_{ik}$?

In order to answer this question we first compute the divergence of $\vec{\bar\sigma}'$, as
\begin{equation}
\begin{aligned}
  (\text{Div}\vec{\bar\sigma}')_1 &= \eta \dfrac{\partial^3 u}{\partial z^2 \partial t}
  + (\zeta+\tfrac43 \eta)\dfrac{\partial}{\partial z}\left[ \left(\dfrac{\partial^2 u}{\partial z\partial t} \right)\left(\dfrac{\partial u}{\partial z}\right)^2\right], \\
    (\text{Div}\vec{\bar\sigma}')_2 &= 0, \\
    (\text{Div}\vec{\bar\sigma}')_3 &= (\zeta+\tfrac43\eta)\dfrac{\partial }{\partial z}\left(\dfrac{\partial u}{\partial z}\dfrac{\partial^2 u}{\partial z\partial t}\right).
\end{aligned}
\end{equation}
Here we remark that the first and last of these components lead to two non-trivial equations of motion, in contrast to the situation encountered in Ref.~ \cite{Zabolotskaya2004}, where there was only one equation.
The second equation here may however be made to disappear if we were to consider soft solids to be perfectly incompressible, and had thus to introduce an arbitrary Lagrange multiplier (for instance, see Ref.~\cite{DeOg10} for a rational inclusion of perfect incompressibility in weakly nonlinear elasticity, and Ref.~\cite{DeSa05} for a derivation of the equations of motion in compressible and incompressible materials and the possible decoupling of longitudinal from transverse waves.)

Next,  we readily derive the \emph{wave equation}
\begin{equation}\label{seq}
     \rho_0 \dfrac{\partial^2 u}{\partial t^2}=\underbrace{\mu  \dfrac{\partial^2 u}{\partial z^2}+\gamma  \dfrac{\partial}{\partial z}\left(\dfrac{\partial u}{\partial z}\right)^3}_{\text{elastic}} 
    + \underbrace{\eta \dfrac{\partial^3 u}{\partial z^2 \partial t}
    +  (\zeta+\tfrac43 \eta)\dfrac{\partial}{\partial z}\left[ \left(\dfrac{\partial^2 u}{\partial z\partial t} \right)\left(\dfrac{\partial u}{\partial z}\right)^2\right]}_{\text{viscous}}.
\end{equation}
The viscous part of this equation is \emph{remarkably} more complex than the simpler term $\eta \partial^3 u/\partial z^2 \partial t$  shown in Eq.~\eqref{viscous.ZH}, which is Eq.~(36) of Ref.~\cite{Zabolotskaya2004}.
Notice that  the last addendum in Eq.~\eqref{seq} is of third order in the strain ${\partial u}/{\partial z}$, exactly as the nonlinear elastic term $\partial/\partial z \left( \partial u / \partial z\right)^3$. Thus, since Eq.~\eqref{equation.motion.ZH} was obtained in 
Ref.\cite{Zabolotskaya2004} by neglecting terms of order \emph{higher} than three, it is our opinion that, in principle, the additional viscous term in \eqref{seq} should be kept and considered for a coherent discussion of wave propagation.
\color{black}
Dimensional analysis reveals that this term would be negligible if fifth-order elastic constants were much larger than lower-order constants. Although this might be the case for some solids, it can be shown by following the steps presented in Ref.\cite{DeOg10} and Ref.~\cite{Ogde74} that for almost incompressible solids, all elastic constants are of the same order as $\mu$ (see Ref.\cite{CaGF03} for experimental evidence).
\color{black}

To illustrate the potential influence of the extra term in Eq.\eqref{seq}, we focus on a staple of acoustic nonlinearity: the finite amplitude, \emph{traveling kink} solution. First we rewrite the equation of motion for the \emph{strain}\cite{DeSa06}: $w \equiv \partial u/\partial z$, as
\begin{equation}
    \rho_0 \dfrac{\partial^2 w}{\partial t^2} = \mu  \dfrac{\partial^2 w}{\partial z^2}
     + \gamma  \dfrac{\partial^2}{\partial z^2}\left(w\right)^3 
    + \eta \dfrac{\partial^3 w}{\partial z^2 \partial t}
    +  \left(\zeta+\tfrac43 \eta \right)\dfrac{\partial^2}{\partial z^2} \left(w^2 \dfrac{\partial w}{\partial t} \right).
\end{equation}
Then we perform the following changes of variables and of function:
\begin{equation}
\xi = \dfrac{\sqrt{\rho_0 \mu}}{\eta} z, \qquad \tau=\dfrac{\mu}{\eta}t, \qquad W = \sqrt{\dfrac{\gamma}{\mu}}w,
\end{equation}
to obtain
\begin{equation}
     \dfrac{\partial^2 W}{\partial \tau^2} = \dfrac{\partial^2 W}{\partial \xi^2}
     + \dfrac{\partial^2}{\partial \xi^2}\left(W\right)^3
    +  \dfrac{\partial^3 W}{\partial \xi^2 \partial \tau}
    +  \dfrac{\mu}{\gamma}\left(\zeta+\tfrac43 \eta \right)\dfrac{\partial^2}{\partial \xi^2} \left(W^2 \dfrac{\partial W}{\partial \tau} \right).
\end{equation}
Looking for a traveling wave solution in the form
\begin{equation}
W(\xi,\tau) = \sqrt{c^2-1}\ \omega(x), \qquad \text{where } x = (c^2-1)(\tau - \xi/c),
\end{equation}
and $c>1$ is the arbitrary, non-dimensional speed, we find the following equation for the amplitude $\omega$,
\begin{equation}
\omega'' = \left(\omega^3\right)'' + \omega''' + \alpha\left(\omega^2 \omega'\right)'', 
\qquad \text{where } \alpha \equiv (\mu/\gamma)\left(\zeta + \tfrac{4}{3}\eta\right)\left(c^2-1\right).
\end{equation}
It can be integrated twice for a finite kink with tails such that $\omega(-\infty)=0$,  $\omega(\infty)=1$,  $\omega'(\pm\infty)= \omega''(\pm\infty)=0$,  to give
\begin{equation}
\omega \left(1-\omega^2\right) = (1 + \alpha \ \omega^2) \omega'.
\end{equation}
Now the variables of this equation can be separated, and by centring the kink so that $\omega(0)=1/2$, we obtain its inverse definition,
\begin{equation}
x = -\ln \dfrac{1}{2\omega}\left[\tfrac{4}{3}(1-\omega^2)\right]^{\frac{\alpha+1}{2}}.
\end{equation}
It is a simple matter to construct the $\omega-x$ curves and then to invert them to generate the $x-\omega$ curves. 
Explicit inversions include 
\begin{equation}
\omega(x) =  \dfrac{1}{\sqrt{1+3\text e^{-2x}}}, \qquad \sqrt{1+\left(\dfrac{3}{4}\text e^x\right)^2} - \dfrac{3}{4}\text e^{-x},
\end{equation}
at $\alpha=0, 1$, respectively.
At $\alpha =0$, there is no extra term in  Eq.\eqref{seq}; as soon as $\alpha \ne 0$, the difference between the incorrect formulation and the proper formulation of viscous effects is felt, as illustrated by Figure 1.

\begin{figure}
\begin{center} 
\epsfig{figure=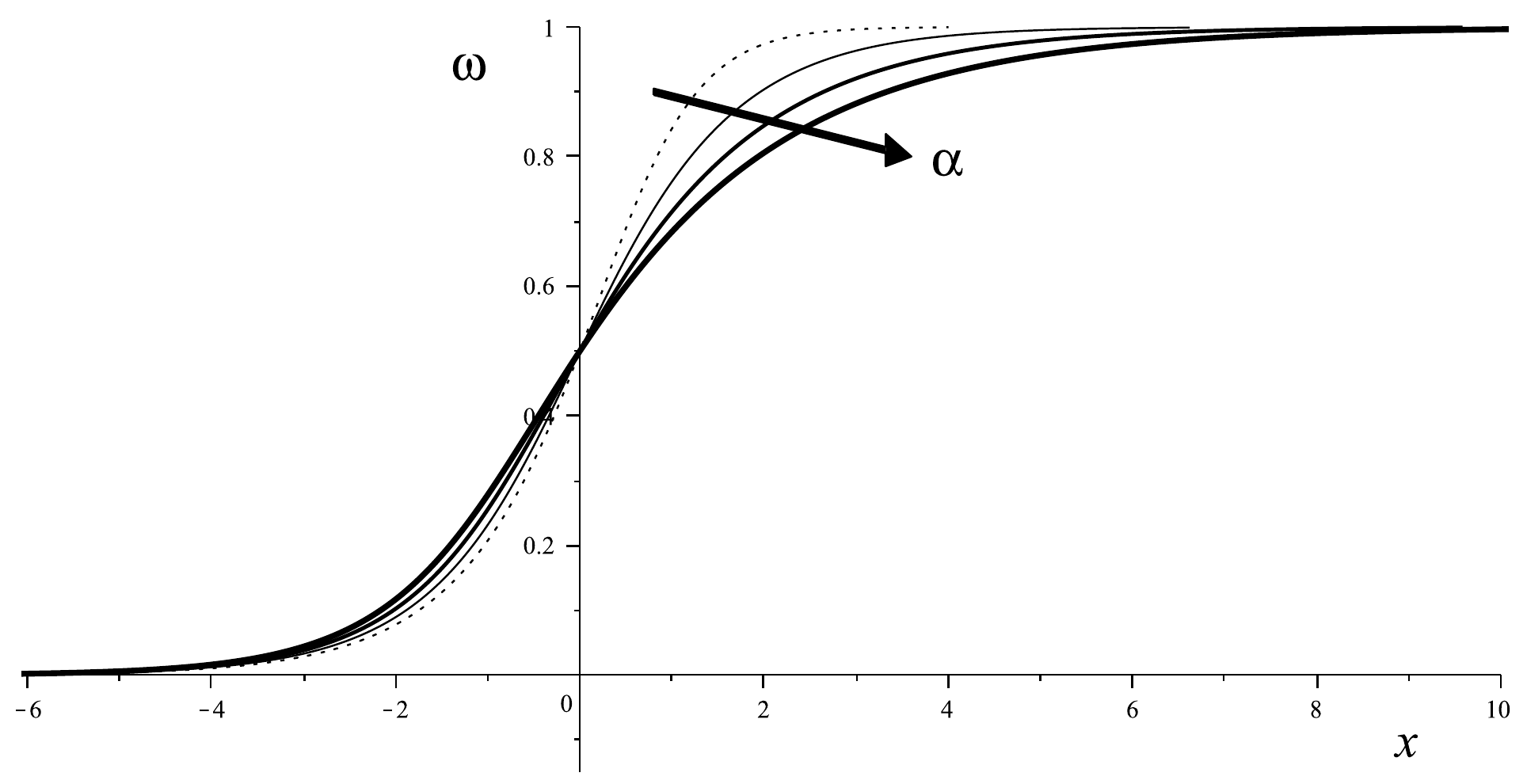, width=0.4\textwidth}
\caption{Finite amplitude travelling transverse strain kink in a viscous soft solid.
The equation of motion has been fully non-dimensionalized. 
The parameter $\alpha$ is a measure of the consequence of properly incorporating viscous effects in the formulation of the equations of motion. At $\alpha=0$ (dotted curve), the viscous formulation does not allow for the principle of objectivity to be respected. As $\alpha$ grows (here, $\alpha=1,2,3$ in turn), the wave front displays a gentler  slope and the wave is more attenuated.
}
\label{fig1}
\end{center}
\end{figure}


\section{Concluding remarks}


Viscous stresses are often introduced in the formulation of nonlinear wave motion problems in order to prevent the formation of shocks, because they confer a parabolic character to the equations of motion for continua. However, as  pointed out earlier by Antman \cite{Ant}, the choices made historically in the
 literature for these stresses sometimes turn out to be physically
unacceptable because their material responses are affected by rigid motions. Here we showed that simply extending the linear Kelvin-Voigt model of differential visco-elasticity from linearized to finite elasticity is not a straightforward process. In particular, one choice is to take the strain-rate effects to be described by the time-derivative of the full Lagrangian strain instead of the infinitesimal strain. With that choice must come great care in formulating a corresponding viscous stress which obeys the fundamental principle of objectivity in the same manner that the elastic stress does. We saw here that this compatibility can be achieved by pre-multiplying the linear expansion of the viscous stress in terms of $\vec{\dot u}$ by the deformation gradient $\vec F$, which complicates greatly the equations of motion, as illustrated in Section \ref{section2}.

An alternative constitutive assumption for modeling viscous effects is to take the Cauchy stress tensor to be linear in $\vec d$, the Eulerian stretching tensor defined in Eq.~\eqref{def.d.lj}. 
That assumption is perfectly coherent with the fundamental principles of mechanics, including frame-invariance, and is aligned with modeling of viscous effects in Fluid Mechanics and the emergence of the Navier-Stokes equations. It has also been used to model nonlinear wave propagation in solids\cite{DeSa05, DeSa06, Jordan}. It does not complicate the equations of motion excessively.

\color{black}
\section*{Acknowledgements}
We are grateful to the Istituto Nazionale di Alta Matematica (Italy) for support through the \emph{PRIN 2009 Matematica e meccanica dei sistemi biologici e dei tessuti molli} for the second author and a short-term Visiting Professor position for the first author.
We thank the anonymous referees for their constructive comments.
\color{black}


\end{document}